\documentclass[aoas,preprint]{imsart}

\RequirePackage[OT1]{fontenc}
\RequirePackage{amsthm,amsmath,graphicx,psfrag,epsf,booktabs,caption,enumerate,color,rotating,multirow,mathtools}
\RequirePackage[authoryear,sectionbib,round]{natbib}
\RequirePackage[colorlinks,citecolor=blue,urlcolor=blue]{hyperref}
\arxiv{arXiv:0000.0000}

\startlocaldefs
\numberwithin{equation}{section}
\theoremstyle{plain}

\endlocaldefs

\begin{document}
	\bibliographystyle{imsart-nameyear}
\begin{frontmatter}
\title{Spline Analysis of Biomarker Data Pooled From Multiple Matched/Nested Case-Control Studies }
\runtitle{Spline Analysis of Biomarker Data\ldots}

\begin{aug}
\author{\fnms{Yujie} \snm{Wu}\thanksref{m1}\ead[label=e1]{yujiewu@hsph.harvard.edu}},
\author{\fnms{Mitchell} \snm{H. Gail}\thanksref{m2}\ead[label=e2]{gailm@mail.nih.gov}},
\author{\fnms{Stephanie} \snm{A. Smith-Warner}\thanksref{m1}\ead[label=e3]{swarner@hsph.harvard.edu}},
\author{\fnms{Regina} \snm{G. Ziegler}\thanksref{m2}\ead[label=e4]{zieglerr@mail.nih.gov}}
\and
\author{\fnms{Molin} \snm{Wang}\thanksref{m1,t1}\ead[label=e5]{stmow@channing.harvard.edu}}
\thankstext{t1}{Corresponding author}
\runauthor{Y. Wu et al.}

\affiliation{Harvard University\thanksmark{m1} 	and National Institutes of Health\thanksmark{m2} }

\address{Y. Wu\\
Department of Biostatistics\\
 Harvard T.H. Chan School of Public Health\\
	Boston, Massachusetts 02115\\
\printead{e1}}

\address{M. H. Gail\\
	Division of Cancer Epidemiology and Genetics\\
	National Cancer Institute, National Institutes of Health\\
	Bethesda, Maryland 20850\\
	\printead{e2}}

\address{S. A. Smith-Warner\\
	Department of Nutrition And\\
	Department of Epidemiology\\
 Harvard T.H. Chan School of Public Health\\
Boston, Massachusetts 02115\\
	\printead{e3}}

\address{R. G. Ziegler\\
	Division of Cancer Epidemiology and Genetics\\
National Cancer Institute, National Institutes of Health\\
Bethesda, Maryland 20850\\
	\printead{e4}}

\address{M. Wang\\
Channing Division of Network Medicine\\
Harvard Medical School And\\
Brigham And Women's Hospital\\
180 Longwood Avenue\\
and\\
Department of Epidemiology And Biostatistics\\
 Harvard T.H. Chan School of Public Health\\
 	Boston, Massachusetts 02115\\
\printead{e5}}
\end{aug}

\begin{abstract}
Pooling biomarker data across multiple studies enables researchers to get more precise estimates of the association between biomarker exposure measurements and disease risks due to increased sample sizes. However, biomarker measurements vary significantly across different assays and laboratories, and therefore calibration of the local laboratory measurements to a reference laboratory is necessary before pooling data. We propose two methods that can estimate a nonlinear relationship between biomarker exposure measurements and disease risks using spline functions with a nested case-control study design: full calibration and internalized calibration. The full calibration method calibrates all observations using a study-specific calibration model while the internalized calibration method only calibrates observations that do not have reference laboratory measurements available. We compare the two methods with a naive method whereby data are pooled without calibration. We find that: (1) Internalized and full calibration methods have substantially better performance than the naive method in terms of average relative bias and coverage rate. (2) Full calibration is more robust than internalized calibration when the size of calibration subsets varies. We apply our methods to a pooling project with nested case-control study design to estimate the association of circulating Vitamin D levels with the risk of colorectal cancer.
\end{abstract}


\end{frontmatter}

\section{Introduction}

It is common to combine biomarker data across different studies to evaluate the association between biomarker exposures and disease risks. The attraction of pooling data for statistical analysis is that the increased sample size from pooling enables us to get more precise estimates of the association between biomarker measurements and disease risks, facilitating subgroup analysis, for example [\citet{ART1,ART2,ART3}]. Studies that pooled data together to examine biomarker-disease associations include: Circulating Vitamin D and Colorectal Cancer [\citet{ART4}], Circulating 25-Hydroxyvitamin D and the Risk of Rarer Cancers [\citet{ART5}], The Endogenous Hormones, Nutritional Biomarkers and
Prostate Cancer Collaborative Group [\citet{ART6,ART7}] and the NCI Breast and Prostate Cancer Cohort Consortium [\citet{ART8}].

We need to account for potential between-study variation of biomarker measurements stemming from assay or laboratory differences before pooling biomarker data. For instance, the variation of measurements of serum 25-hydroxyvitamin D (25(OH)D) concentration can reach up to 40\% across different assays, laboratories and even seasons of a year [\citet{ART9,ART10,ART11}]. Also, hormones such as estradiol and testosterone have high variation across assays and laboratories [\citet{ART1,ART2,ART3}]. Therefore, additional steps are needed to standardize biomarker measurements across different laboratories before data analysis. We employ a study-specific calibration model for each study to transform the original data into comparable measurements. The calibration is performed by first selecting a reference laboratory, to which a subset of biospecimens from each study are sent for re-assaying. We can therefore estimate a calibration model for each study based on the local and corresponding reference laboratory measurements among the subset of biospecimens selected for re-assaying. The models are then used to impute the reference laboratory measurements for the remaining observations with only local laboratory measurements available. Due to their rarity, cases are not usually used for assay calibration. Instead, controls are usually chosen for re-assaying in a reference laboratory [\citet{ART12}].

The calibration approach can also be regarded as a covariate measurement error problem. Instead of observing the true covariate measurement $X$ for each subject, we only have surrogate measurement $W$ available[\citet{ART13}]. The study-specifc calibration model is therefore trying to correct for the measurement errors. In our paper, we build our method based on the regression calibration approach which is widely used for correcting for measurement error-caused biases[\citet{ART13, ART22}].

\citet{ART15} proposed methods for pooling biomarker data from nested case-control studies. 
However, their method assumes that bioma-rker measurements have a linear relationship with log relative risk(RR) of diseases where a nonlinear dose-response relationship of biomarker measurements and disease risk is often observed in practice. For example, a nonlinear inverse association between 25(OH)D and breast cancer risk was observed among postmenopausal women [\citet{ART16}]; similarly, a nonlinear relationship between 25(OH)D and indices of arterial stiffness and arteriosclerosis was found in an elderly population in the Netherlands [\citet{ART17}]. Thus, restricting the biomaker-disease relationship to be linear may lead to biased estimates when the true association is nonlinear. In this paper, we extend the method for pooling nested case-control studies to allow for a nonlinear relationship between biomarker effects and disease risks using spline functions [\citet{ART18}]. We first calibrate biomarker measurements across multiple studies using the Internalized and Full calibration methods [\citet{ART12}]. We then obtain estimates for the coefficients of the spline functions based on an approximate likelihood function. However, the variances of the estimates are not directly obtainable due to the uncertainty in the estimation of calibration parameters, and therefore we use empirical sandwich variance estimates.

In Section 2, we present the models and statistical methods. In Section 3, we evaluate the performance of our methods in simulation studies.  In Section 4, we apply the methods to a pooling project investigating the relationship between circulating Vitamin D levels with colorectal cancer, and we present a discussion in Section 5.

\section{Methods}
\subsection{Model and approximate conditional likelihood}

Let $1,\ldots, S$ index the studies that contributed to the pooling project, where the first $Q$ studies use local laboratories for measurement of biomarkers. We use $n_{sj}$ and $ m_{sj}$ to denote the number of cases and controls, respectively in the $j^{\text{th}}$ stratum of the $s^{\text{th}}$ study. Within each stratum, we use $i$ to index individuals, where controls correspond to $i=1, 2,\ldots, m_{sj}$, and the cases correspond to $i=m_{sj}+1, m_{sj}+2, \ldots, m_{sj}+n_{sj}$.

Let $Y_{sji}$ denote the binary disease status of each individual, $X_{sji}$ be the biomarker measurements taken from the reference laboratory, $W_{sji}$ be the biomarker measurements taken from local laboratories, and let $\mathbf{Z}_{sji}$ be a vector of potential confounders. For each study using a local laboratory, a subset of samples were sent to the reference laboratory for re-assaying to obtain reference biomarker measurements $X_{sji}$, and we refer this subset as the calibration subset. Therefore, for studies that use local laboratories for biomarker measurements, $X_{sji}$ are available only in the calibration subsets, and $W_{sji}$ are available for all individuals. Since the local measurements can vary systematically across different studies, using $W_{sji}$ instead of $X_{sji}$ can lead to significantly biased estimates of the biomarker-disease relationship. We propose to calibrate the local laboratory measurements by building calibration models based on the calibraton subsets in each local laboratory. 

To model the possibly nonlinear relationship between the biomarker and the disease risk, we use spline functions [\citet{ART18}]. Let $s$ index a study and $j$ a stratum within a study. The logistic regression model with spline functions can be written as
\begin{equation*}
\text{logit}(P(Y_{sji}=1|X_{sji}, \mathbf{Z}_{sji}))=\beta_{0sj}+\boldsymbol{\beta}^T_{X}\boldsymbol{f}(X_{sji})+\boldsymbol{\beta}_{\mathbf{Z}}^T\mathbf{Z}_{sji}\,\, ,
\end{equation*}
where $\beta_{0sj}$ is the study and stratum-specific intercept, and $\boldsymbol{f}(X_{sji})$ is a $K\times 1$ column vector of spline basis functions at $X_{sji}$, $K$ depends on the type of spline functions and the number of knots selected, $\boldsymbol{Z}_{sji}$ is a $P\times 1$ column vector of potential confounders, and $\boldsymbol{\beta}_X$ and $\boldsymbol{\beta}_\mathbf{Z}$ are column vectors of the corresponding regression coefficients. Note that in nested case-control studies with density sampling [\cite{ART14}] $\boldsymbol{\beta}_X^T\left[\boldsymbol{f}(X_{sji})-\boldsymbol{f}(X_{sji'})\right]$ represents the log relative risk(RR) associated with two distinct biomarker measurements $X_{sji}$ and $X_{sji'}$ in the same stratum. We focus on point and interval estimates of $\boldsymbol{\beta}_X$ in this paper. 

Let vectors $\boldsymbol{X}_{sj}$, $\boldsymbol{W}_{sj}$ and matrix $\boldsymbol{Z}_{sj}$ contains the corresponding measurements of individuals from the $j^{\text{th}}$ stratum of the $s^{\text{th}}$ study. The conditional likelihood function is [\citet{ART14}]:
\begin{equation*}
\tiny
\begin{split}
L&=\prod_{s}\prod_{j}P( Y_{sj1}=0,\ldots,Y_{sjm_{sj}}=0,Y_{sj(m_{sj}+1)}=1,\ldots,Y_{sj(m_{sj}+n_{sj})}=1|\boldsymbol{X}_{sj}, \boldsymbol{Z}_{sj}, \sum_{i=1}^{m_{sj}+n_{sj}}Y_{sji}=n_{sj}    )\\
&=\prod_s\prod_j\frac{ \prod_{l=1}^{n_{sj}}\exp\left(\beta_{0sj}+\boldsymbol{\beta}_X^T\boldsymbol{f}(X_{sj(m_{sj}+l)})+\boldsymbol{\beta}_{\mathbf{Z}}^T\mathbf{Z}_{sj(m_{sj}+l)}  \right)    }{  \sum_{(i_1,\ldots,i_{n_{sj}})\in A}\prod_{l=1}^{n_{sj}} \exp(\beta_{0sj}+\boldsymbol{\beta}_X^T\boldsymbol{f}(X_{sji_l})+\boldsymbol{\beta}_{\mathbf{Z}}^T\mathbf{Z}_{sji_l}  )      }\\
&=\prod_s\prod_j\left( 1+\sum_{(i_1,\ldots,i_{n_{sj}})\in A'}\exp\left(  \boldsymbol{\beta}_X^T\sum_{l=1}^{n_{sj}}\left[ \boldsymbol{f}(X_{sji_l})-\boldsymbol{f}(X_{sj(m_{sj}+l)})\right]  +\boldsymbol{\beta}_\mathbf{Z}^T\sum_{l=1}^{n_{sj}}\left[ \mathbf {Z}_{sji_l}-\mathbf{Z}_{sj(m_{sj}+l)}  \right]  \right)    \right)^{-1},
\end{split}
\end{equation*}
where $A$ is the set of all subsets of indices of size $n_{sj}$ of the set $\{1, 2, \ldots, m_{sj},\\ m_{sj}+1, \ldots, m_{sj}+n_{sj}\}$ and $(i_1, i_2, \ldots, i_{n_{sj}})$ corresponds to one specific such subset of size $n_{sj}$. $A'$ is the subset of $A$ that excludes subset where $i_1=m_{sj}+1, i_2=m_{sj}+2, \ldots, i_{n_{sj}}=m_{sj}+n_{sj}$.

The conditional likelihood function cannot be calculated directly since $\boldsymbol{X}_{sj}$ is not available for all individuals if the biomarker measurements were taken only in a local laboratory. To derive an approximate conditional likelihood under the nested case-control study design, we make the `surrogacy' assumption that takes account of the study design:
\begin{equation*}
f(\mathbf{Y}_{sj}|\mathbf{X}_{sj}, \mathbf{W}_{sj}, \mathbf{Z}_{sj},\sum_{i=1}^{n_{sj}+m_{sj}}Y_{sji}=n_{sj})=f(\boldsymbol{Y}_{sj}|\boldsymbol{X}_{sj}, \mathbf{Z}_{sj},\sum_{i=1}^{n_{sj}+m_{sj}}Y_{sji}=n_{sj})
\end{equation*}
which states that the outcome is conditionally independent of $\mathbf{W}_{sj}$ given the reference laboratory measurements, other covariates of interest and the design schema.

Under this surrogacy assumption, the likelihood contributed from a stratum using local laboratory biomarker measurements is:
\begin{equation*}
\tiny
\begin{split}
L_{sj}=&P\left(Y_{sji}=0,\ldots, Y_{sjm_{sj}}=0,Y_{sj(m_{sj}+1)}=1,\ldots, Y_{sj(m_{sj}+n_{sj})}=1|\boldsymbol{W}_{sj}, \boldsymbol{Z}_{sj},\sum_{i=1}^{m_{sj}+n_{sj}}Y_{sji}=n_{sj}\right)\\
=&\int P\left(Y_{sji}=0,\ldots, Y_{sjm_{sj}}=0,Y_{sj(m_{sj}+1)}=1,\ldots, Y_{sj(m_{sj}+n_{sj})}=1|\boldsymbol{X}_{sj},\boldsymbol{W}_{sj}, \boldsymbol{Z}_{sj},\sum_{i=1}^{m_{sj}+n_{sj}}Y_{sji}=n_{sj}\right)\\
&\times P\left(\boldsymbol{X}_{sj}|\boldsymbol{W}_{sj},\boldsymbol{Z}_{sj},\sum_{i=1}^{m_{sj}+n_{sj}}Y_{sji}=n_{sj}\right)\,d\boldsymbol{X}_{sj}\\
=&\int \left( 1+\sum_{(i_1,\ldots,i_{n_{sj}})\in A'}\exp\left(  \boldsymbol{\beta}_X^T\sum_{l=1}^{n_{sj}}\left[ \boldsymbol{f}(X_{sji_l})-\boldsymbol{f}(X_{sj(m_{sj}+l)})\right]  +\boldsymbol{\beta}_\mathbf{Z}^T\sum_{l=1}^{n_{sj}}\left[ \mathbf{Z}_{sji_l}-\mathbf{Z}_{sj(m_{sj}+l)}  \right]  \right)    \right)^{-1}\\
&\times P\left(\boldsymbol{X}_{sj}|\boldsymbol{W}_{sj},\boldsymbol{Z}_{sj},\sum_{i=1}^{m_{sj}+n_{sj}}Y_{sji}=n_{sj}\right)\,d\boldsymbol{X}_{sj}\\
=&E_{ \boldsymbol{X}_{sj}|\boldsymbol{W}_{sj},\boldsymbol{Z}_{sj},\sum_{i=1}^{n_{sj}+m_{sj}}Y_{sji}=n_{sj} }  \Bigg(\Bigg( 1+\sum_{(i_1,\ldots,i_{n_{sj}})\in A'}\exp\Bigg(  \boldsymbol{\beta}_X^T\sum_{l=1}^{n_{sj}}\left[ \boldsymbol{f}(X_{sji_l})-\boldsymbol{f}(X_{sj(m_{sj}+l)})\right]  \\
&+\boldsymbol{\beta}_\mathbf{Z}^T\sum_{l=1}^{n_{sj}}\left[ \mathbf{Z}_{sji_l}-\mathbf{Z}_{sj(m_{sj}+l)}  \right]  \Bigg)    \Bigg)^{-1}   \Bigg)
\end{split}
\end{equation*}
where the surrogacy assumption justifies the third equation.

Let $\tiny{F= \left( 1+\sum_{(i_1,\ldots,i_{n_{sj}})\in A'}\exp\left(  \boldsymbol{\beta}_X^T\sum_{l=1}^{n_{sj}}\left[ \boldsymbol{f}(X_{sji_l})-\boldsymbol{f}(X_{sj(m_{sj}+l)})\right]  +\boldsymbol{\beta}_\mathbf{Z}^T\sum_{l=1}^{n_{sj}}\left[ \mathbf{Z}_{sji_l}-\mathbf{Z}_{sj(m_{sj}+l)}  \right]  \right)    \right)^{-1}} .$ We expand $F$ in Taylor series around $\widetilde{\boldsymbol{X}}_{sj}=E\left( \boldsymbol{X}_{sj}|\boldsymbol{W}_{sj},\boldsymbol{Z}_{sj},\sum_{i=1}^{m_{sj}+n_{sj}}Y_{sji}=n_{sj}   \right)$, yielding the following approximate likelihood contribution from the $j^{\text{th}}$ stratum of the $s^{\text{th}}$ study:
\begin{equation*}
\scriptsize
\widetilde{L}_{sj}= \left( 1+\sum_{(i_1,\ldots,i_{n_{sj}})\in A'}\exp\left(  \boldsymbol{\beta}_X^T\sum_{l=1}^{n_{sj}}\left[ \boldsymbol{f}(\widetilde{X}_{sji_l})-\boldsymbol{f}(\widetilde{X}_{sj(m_{sj}+l)})\right]  +\boldsymbol{\beta}_\mathbf{Z}^T\sum_{l=1}^{n_{sj}}\left[ \mathbf{Z}_{sji_l}-\mathbf{Z}_{sj(m_{sj}+l)}  \right]  \right)    \right)^{-1}
\end{equation*}
The approximation performs best when the conditional variance and covariance of $\boldsymbol{X}_{sj}$ are small or the biomarker effect is not strong. Section 1 of Supplementary Materials provides a detailed derivation of the approximate conditional likelihood and the conditions when the approximation works well. 
To get an estimate of $\widetilde{X}_{sji}$, we propose the calibration model in the next section.

\subsection{Calibration model}
In each study that uses local laboratories, the study-specific calibration models can be developed by utilizing the subset of controls selected for re-assaying in the reference laboratory which therefore have biomarker measurements from both local and reference laboratories. The calibration models can thus be used to impute the reference biomarker measurements for the remaining subjects that only have local laboratory measurements.

We make the calibration assumption that the calibration model will not be heavily affected by other covariates apart from the local laboratory measurement for the same participant; that is, $$\small{E\left(X_{sji}|\boldsymbol{W}_{sj}, \boldsymbol{Z}_{sj},\sum_{i=1}^{m_{sj}+n_{sj}}Y_{sji}=n_{sj} \right)\approx  E\left(X_{sji}|W_{sji},\sum_{i=1}^{m_{sj}+n_{sj}}Y_{sji}=n_{sj} \right)}$$ 
We assume a linear relationship between reference laboratory measurements and local laboratory measurements for calibration.
\begin{equation}
E\left(X_{sji}|W_{sji},\sum_{i=1}^{m_{sj}+n_{sj}}Y_{sji}=n_{sj} \right)=a_s+b_sW_{sji},\label{eq1}
\end{equation}
where $a_s,b_s$ are study-specific calibration model coefficients. Notice that the calibration parameters are the same across different strata in each study. However, we can relax this constraint by assuming the calibration models also depend on matching factors:
$E\left(X_{sji}|W_{sji},\boldsymbol{M}_{sji},\sum_{i=1}^{m_{sj}+n_{sj}}Y_{sji}=n_{sj} \right)=a_s+b_sW_{sji}+\boldsymbol{c}_s^T\boldsymbol{M}_{sji}$, where $\boldsymbol{M}_{sji}$ are the matching factors that may have an effect on the calibration models. \citet{ART12} suggested that the equation \ref{eq1} is sufficient in most study settings. In \ref{eq1}, although a linear term of $W_{sji}$ is typically sufficient to model the $X_{sji}-W_{sji}$ relationship, nonlinear terms in $W_{sji}$ can also be included if appropriate. 

The calibration models are usually fitted among controls because case bio-specimens are often not available, and therefore the calibration model used in practice is:
\begin{equation*}
E\left(X_{sji}|W_{sji},Y_{sji}=0\right)=a_{s,co}+b_{s,co}W_{sji}
\label{calib}
\end{equation*}
where we use $a_{s,co}, b_{s,co}$ to denote the calibration models fitted among controls only. $\widehat{a}_{s,co}, \widehat{b}_{s,co}$ are generally not consistent estimates of $a_s, b_s$. \citet{ART15} gave conditions for $a_{s,co}\approx a_s, b_{s,co}\approx b_s$ under bivariate normality of $\boldsymbol{X}_{sj}$ and $\boldsymbol{W}_{sj}$ in a 1:1 nested case-control study. It is straightforward to generalize their results to $n_{sj}:m_{sj}$ matching. This yields $b_{s,co}\approx b_s$ when $Var(\boldsymbol{X}_{sj}|\sum_{i=1}^{m_{sj}+n_{sj}}Y_{sji}=n_{sj})\approx Var(\boldsymbol{X}_{sj}|\boldsymbol{Y}_{sj}=0)$, and $a_{co}\approx a_s$ when $Var(\boldsymbol{X}_{sj}|\sum_{i=1}^{m_{sj}+n_{sj}}Y_{sji}=n_{sj})\approx Var(\boldsymbol{X}_{sj}|\boldsymbol{Y}_{sj}=0)$ and $E(\boldsymbol{X}_{sj}|\sum_{i=1}^{m_{sj}+n_{sj}}Y_{sji}=n_{sj})\approx E(\boldsymbol{X}_{sj}|\boldsymbol{Y}_{sj}=0)$. In addition, if the biomarker effect is small(i.e. $\boldsymbol{\beta}_X\approx 0$), $\widehat{a}_{s,co}, \widehat{b}_{s,co}$ will also be close to $a_s, b_s$. 

Define $\widetilde{X}_{sji}$ as the `reference' biomarker measurement as an alternative to $X_{sji}$ for use in the pooled analysis. If the study already used the reference laboratory for measurement of all participants, no calibration is needed and $\widetilde{X}_{sji}=X_{sji}$. For studies that used local laboratory for measurement, we define the full calibration and internalized calibration methods [\cite{ART12}] as follows:

\begin{equation*}
\begin{split}
\text{Full Calibration}&: \widetilde{X}_{sji}=\widehat{E}(X_{sji}|W_{sji},Y_{sji}=0)\\
\text{Internalized Calibration}&:\\
\widetilde{X}_{sji}=&\begin{cases}
X_{sji}, & \text{if reference lab}\\
&\text{ measurement available}\\
\widehat{E}(X_{sji}|W_{sji},Y_{sji}=0), & \text{otherwise}.
\end{cases}
\end{split}
\end{equation*}

Therefore, for studies using local laboratories for biomarker measurement, all participants' biomarker measurements are calibrated under the full calibration method while under the internalized calibration method, the biomarker measurements are calibrated for participants who only have the local laboratory measurements available.
\subsection{Parameter Estimation}
We define $\boldsymbol{a}=\left[a_1,a_2,\ldots,a_Q\right]$, $\boldsymbol{b}=\left[b_1,b_2,\ldots,b_Q \right]$, and the dose-response parameters $\boldsymbol{\beta}=\left[\boldsymbol{\beta}_X,\boldsymbol{\beta}_{\boldsymbol{Z}} \right]$. The collective set of parameters to be estimated is therefore  $\boldsymbol{\theta}=\left[\boldsymbol{a},\boldsymbol{b},\boldsymbol{\beta} \right]$. The corresponding estimating equations are 
$\left[\boldsymbol{\psi}_{\boldsymbol{a}}, \boldsymbol{\psi}_{\boldsymbol{b}},\boldsymbol{\psi}_{\boldsymbol{\beta}_X},\boldsymbol{\psi}_{\boldsymbol{\beta}_{\boldsymbol{Z}}} \right]=\boldsymbol{0}$, where $\boldsymbol{\psi}_{\boldsymbol{a}}, \boldsymbol{\psi}_{\boldsymbol{b}},\boldsymbol{\psi}_{\boldsymbol{\beta}_X},\boldsymbol{\psi}_{\boldsymbol{\beta}_{\boldsymbol{Z}}}$ are the estimating functions for their corresponding parameters. Section 2 in the supplementary material contains the mathematical details.

We propose to obtain the point estimate for $\boldsymbol{\beta}$ with a two-step pseudo maximum likelihood method [\citet{ART19}]. In the first step, estimates of $\boldsymbol{a},\boldsymbol{b}$ of the calibration models are obtained by fitting linear regressions on the subset of controls chosen for re-assaying in the reference labs, and in the second step, $\boldsymbol{\beta}$ are obtained  using pseudo-maximum conditional likelihood method by solving the estimating equations $\left[\boldsymbol{\psi}_{\boldsymbol{\beta}_X}(\boldsymbol{\widehat{a}},\boldsymbol{\widehat{b}}),\boldsymbol{\psi}_{\boldsymbol{\beta}_{\boldsymbol{Z}}}(\boldsymbol{\widehat{a}},\boldsymbol{\widehat{b}}) \right]=0$, where $\boldsymbol{a},\boldsymbol{b}$ are replaced by their estimates in the previous step.

We use a sandwich variance formula over all the estimating equations to estimate $Var(\widehat{\boldsymbol{\beta}}_X)$. Therefore, $\widehat{Var}(\widehat{\boldsymbol{\beta}}_X)$ can be found by collecting the corresponding diagonal element of the sandwich variance matrix. Section 3 of the supplementary materials contains mathematical details.

\section{Simulations}

We performed simulations for a 1:1 matched case-control study design. Define $\epsilon_{sji}$ as the error term in the linear model of $X_{sji}$ on $W_{sji}$. We assume a similar multivariate normal distribution of $X_{sji}, W_{sji},\epsilon_{sji}$ as \cite{ART15} such that
\begin{equation*}
\small
\left(\begin{array}{c}
X_{sji}\\
W_{sji}\\
\epsilon_{sji}
\end{array}\right)\sim \text{MVN}\left(
\left( \begin{array}{c}
\mu_x  \\
(\mu_x-a_s)/b_s\\
0
\end{array}\right),
\left(
\begin{array}{ccc}
\sigma_x^2&b_s\sigma_{ws}^2&\sigma_x^2-b_s^2\sigma_{ws}^2  \\
b_s\sigma_{ws}^2 &\sigma_{ws}^2&0\\
\sigma_x^2-b_s^2\sigma_{ws}^2&0&\sigma_x^2-b_s^2\sigma_{ws}^2
\end{array}
\right)\right)
\end{equation*}

This distribution yields the calibration model $E(X_{sji}|W_{sji})=a_s+b_sW_{sji}$ and $Cov(W_{sji},\epsilon_{sji})=0$. The data were generated for each stratum of each study first, and then a case and control were randomly chosen in each stratum.  In the simulation, we set $\mu_x=0, \sigma_x^2=1$. We assumed four studies in the pooled analysis with 500 case-control pairs(i.e. 1000 total subjects) in each study, and the calibration parameters were set to be: $
\boldsymbol{a}=\left[ -3,1,-1,3\right]$, and $\boldsymbol{b}=\left[0.5, 0.75, 1.25, 1.5 \right]$. We set $\text{Var}(W_{sji})=\sigma_{ws}^2=\left[3.8, 1.7, 0.6, 0.4 \right]$ to ensure a wide range of variation of local laboratory measurements.  The stratum specific intercept was assumed to follow a normal distribution with mean 0 and variance 0.01: $\beta_{0sj}\sim N(0, 0.01)$.

The spline functions were chosen to be restricted cubic splines with three knots fixed at the $(25^{\text{th}}, 50^{\text{th}}, 75^{\text{th}})$ quantile of $N(0,1)$. We assume a simple risk model without additional covariates:
\begin{equation*}
\text{logit}\left(Y_{sji}=1|X_{sji} \right)=\beta_{0sj}+\beta_{X_1}f_1(X_{sji})+\beta_{X_2}f_2(X_{sji}),
\end{equation*}
where $f_1(X_{sji})=X_{sji}$, and $f_2(X_{sji})=(X_{sji}-t_1)_+^3-(X_{sji}-t_2)_+^3\frac{t_3-t_1}{t_3-t_2}+(X_{sji}-t_3)_+^3\frac{t_2-t_1}{t_3-t_2}$, and $t_1,t_2,t_3$ are the three knots mentioned above [\citet{ART18}]. Note that $\beta_{X_2}=0$ implies a linear relationship between $X$  and $Y$. 

The simulations were performed 1000 times for different combinations of $(\beta_{X_1},\beta_{X_2})$ and calibration proportions at 5\%, 15\%, and 30\%, which is defined as the proportion of controls in each study that are sent for re-assaying in the reference laboratory.

We compare the performance of both the Internalized(IN) and Full(FC) calibration methods in terms of average relative bias($(\widehat{\beta}-\beta)/\beta$) over the simulation replicates and coverage rate, which is defined as the proportion of the estimated 95\% confidence intervals containing the true value. We also included the Naive(N) method for comparison, where no calibration is performed and the conditional logistic regression is fitted using the local laboratory measurements directly.

The simulation results in Tables \ref{point1} and \ref{point2} of the main paper are for a biomarker that has an inverse effect on the diesease risk and Supplementary tables \ref{point1} and \ref{point2} in supplementary materials are for a biomarker that has a positive association with the disease risk. The Naive method performs poorly in all scenarios regardless of the calibration proportions. The average relative bias of the naive estimates are typically larger than 0.3, and the coverage rates are typically below 70\%. The Internalized and Full calibration estimates have consistently better performance than the Naive method. Full calibration estimates are robust over many combinations of coefficients and calibration proportions, where the average relative bias is typically below 0.10 when calibration proportion is 5\% and below 0.05 when calibration proportion is 15\% and 30\%. The coverage rates range from 93\% to 97\% which are close to 95\% nominal level. The internalized calibration estimates are less robust than full calibration estimates, which tend to be more biased when the calibration proportion is large. Section 4 of the supplementary materials contains a mathematical justification for the relative performance of the internalized and full calibration models. 
   
\begin{figure}[!htb]
	\minipage{0.5\textwidth}
	\includegraphics[width=2.3in]{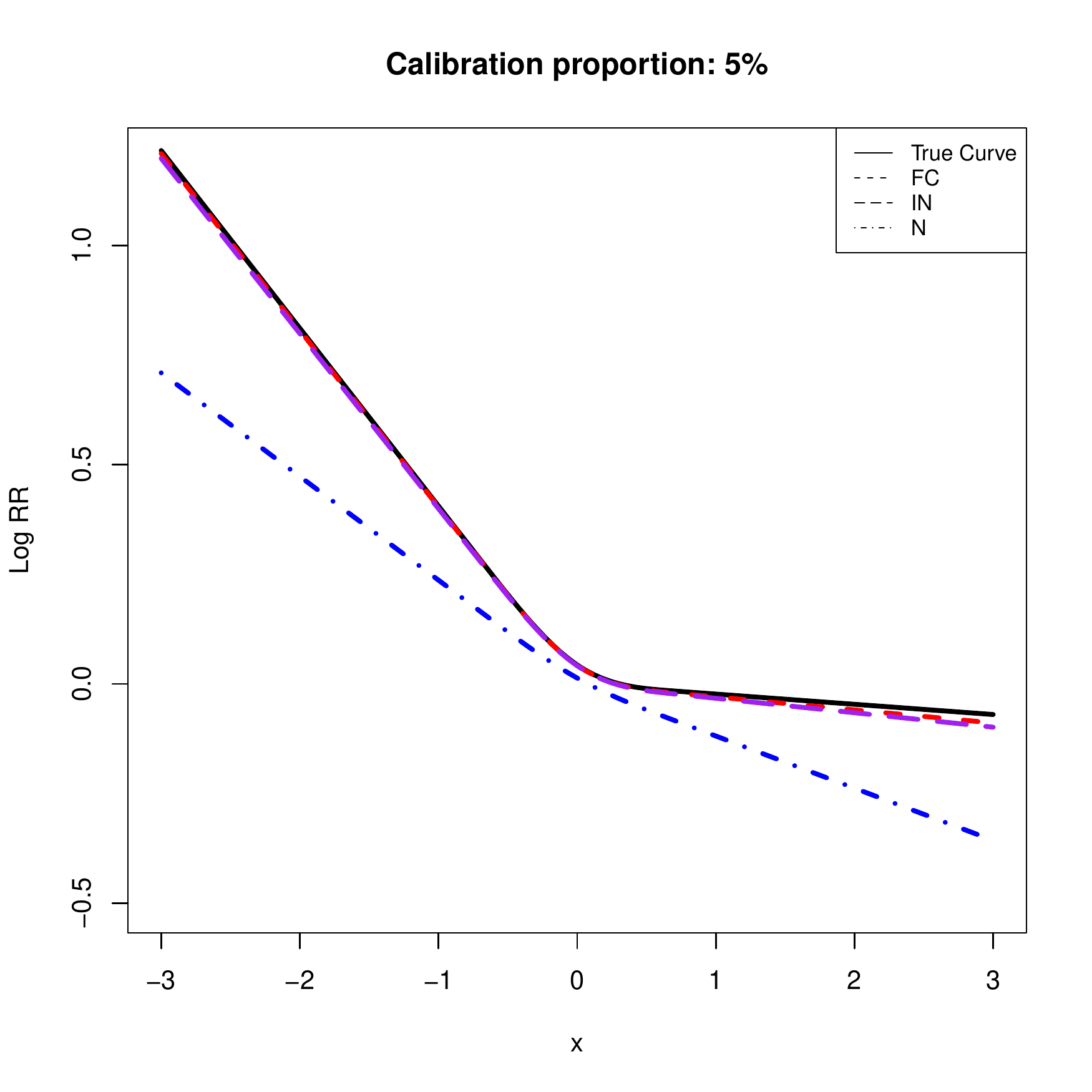}
	\endminipage\hfill
	\minipage{0.5\textwidth}%
	\includegraphics[width=2.3in]{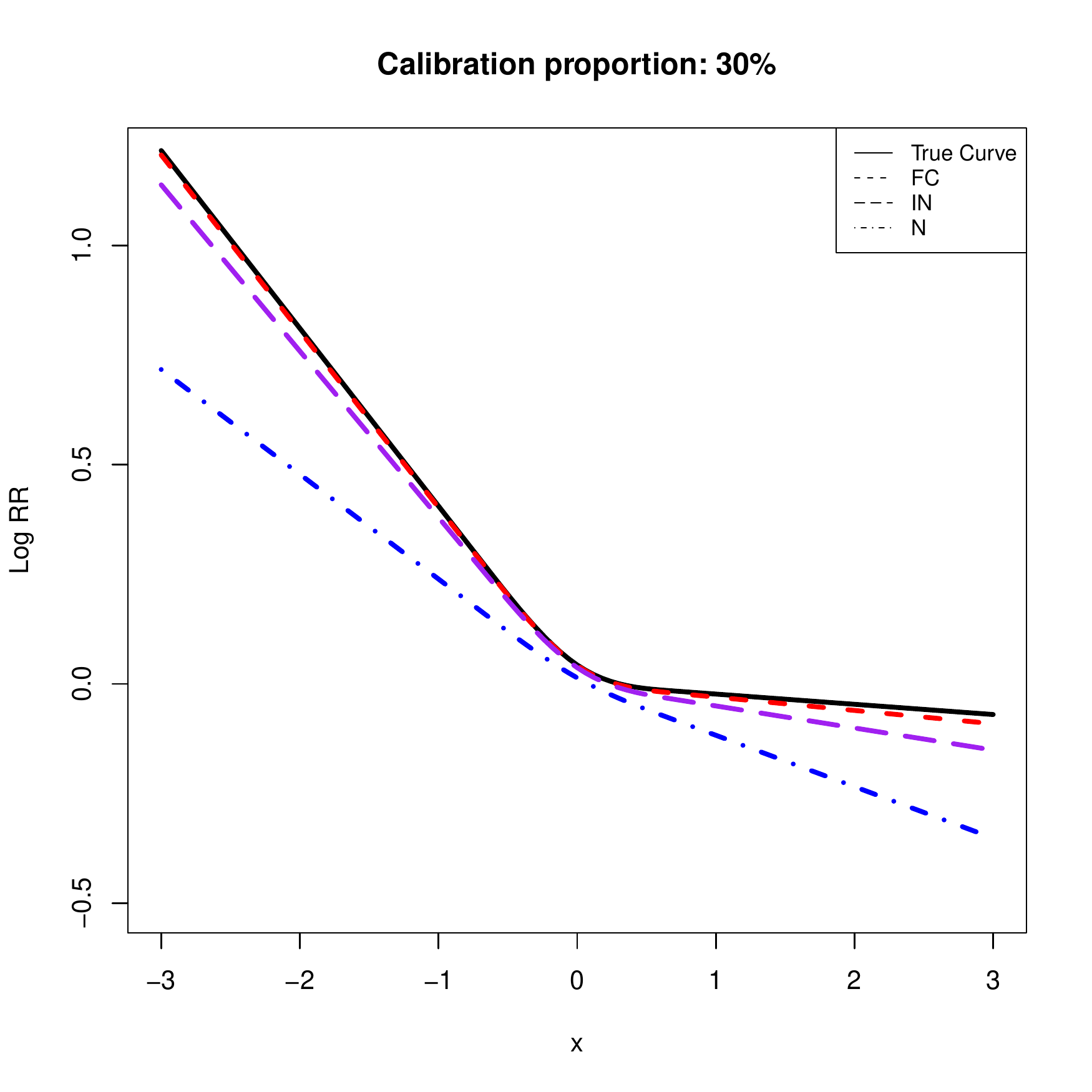}
	\endminipage
	\caption{The curve reflecting the association of biomarker measurements on disease risk, where the x-axis is the biomarker measurement and y-axis is the log RR of the disease. The solid line is the true curve, and the dotted and dashed lines were estimated using internalized(IN) and full calibration methods(FC) respectively, while the dashed-dotted line is estimated using the naive method(N). The calibration proportion is 5\%(left) and 30\%(right), and the coeffcients of the spline functions are set to be $-\text{log}(1.5)\approx-0.41$ and 0.14, respectively.}
	\label{fig1}
\end{figure}

We also plotted the curves reflecting the biomarker-disease association. The x-axis represents the biomarker values, and the y-axis is the log RR. Figure \ref{fig1} is for the scenario where $\beta_{X_1}=-\log(1.5)\approx -0.41, \beta_{X_2}=0.14$, and the calibration proportions were set to be 5\% and 30\% respectively. We can see that the curve estimated using naive method deviates from the true curve substantially, while the curves estimated using the internalized and full calibration methods are closer to the true curve. As the calibration proportion increases, the curve estimated using the full calibration method is closer to the true curve than the internalized calibration method. Supplementary figure \ref{fig1} in the supplementary materials describes the scenario when $\beta_{X_1}=\log(1.75)\approx 0.56, \beta_{X_2}=-0.16$, and the calibration proportion were also set to be 5\%, 30\% respectively. We can see similar behaviors of the three estimated curves, where the full calibration method led to the estimated curve that is closest to the true curve and is robust over all calibration proportions.

In addition, we changed $\sigma_{ws}^2$ in the simulation setup to vary the ratio of $\frac{\sigma_{ws}^2}{\sigma_{X}^2}$. This ratio for each study was chosen from 0.75, 0.85, 0.90, and 0.95. The simulation results in table \ref{sense2} show that the performance of calibration models improves as this variance ratio increases, that is when the error term in the calibration model $X|W$ becomes smaller. The full calibration method is more robust than the internalized method with smaller relative bias and coverage rate closer to the 95\% nominal level for all the variance ratios considered.

\section{Applied Example}
To illustrate, we applied our methods to evaluate the association of circulating Vitamin D level (25(OH)D) with colorectal cancer incidence. We based the example on two large cohort studies in the United States: Nurses' Health Study (NHS) [\citet{ART20}] and Health Professionals Follow-up Study (HPFS) [\citet{ART21}]. In NHS, 121,701 female nurses aged 30 to 55 in 1976 were enrolled while the HPFS had an enrollment of 51,529 male health professionals aged 40 to 75 in 1986. A subset of participants were selected from 1989 to 1995 to obtain the measurements of their biomarkers including 25(OH)D for both studies. Individuals without colorectal cancer outcome or 25(OH)D measurement were excluded from the pooling analysis. In all, our pooling analysis consisted of 1,876 subjects. Twenty-nine controls in each nested case-control study were selected to have their blood samples re-assayed at Heartland Assays, LLC(Ames, IA), the reference laboratory, during 2011-2013 [\citet{ART4}].
 
Table \ref{calibmodel} presents sample sizes of the main studies and parameter estimates along with standard errors of the study-specific calibration models. The potential confounders adjusted for in the conditional logistic regression model included smoking(yes/no), BMI(greater or less than 25), physical activity(continuous) and family history of myocardial infarction(yes/no). We chose a restricted cubic spline with three knots at the 25\%, 50\% and 75\% quantiles of reference 25(OH)D measurements to estimate how log RR representing the Vitamin D-colorectal cancer relationship changes with the Vitamin D level.  Table \ref{clogit} presents the coefficient estimates along with corresponding 95\% confidence intervals.

As shown in Table \ref{clogit}, we obtained similar point and confidence interval estimates of coefficients from the internalized and full calibration methods. We observed a significant linear relationship between 25(OH)D measurements with log RR of colorectal cancer(p-value = 0.0211 and 0.0217, for the internalized and full calibration methods respectively), while the nonlinear relationship between 25(OH)D measurements with log RR of colorectal cancer was not significant(p-value = 0.2162 and 0.2219, for the internalized and full calibration methods respectively). Therefore we concluded that circulating Vitamin D level has a significant linear association the log relative risk of colorectal cancer.

After dropping the nonlinear term from the logistic regression model, in Table \ref{clogit}, the point estimate of biomarker effect on the log RR of coleractal cancer was $-0.0059$($\text{RR}=0.9941$), and the 95\% confidence interval was $(-0.0108, -0.0010)$ with a p-value of 0.0177, suggesting a significant negative linear relationship between levels of circulating 25(OH)D measurements and log RR of colorectal cancer.

\begin{figure}[!htb]
	\includegraphics[width=2.3in]{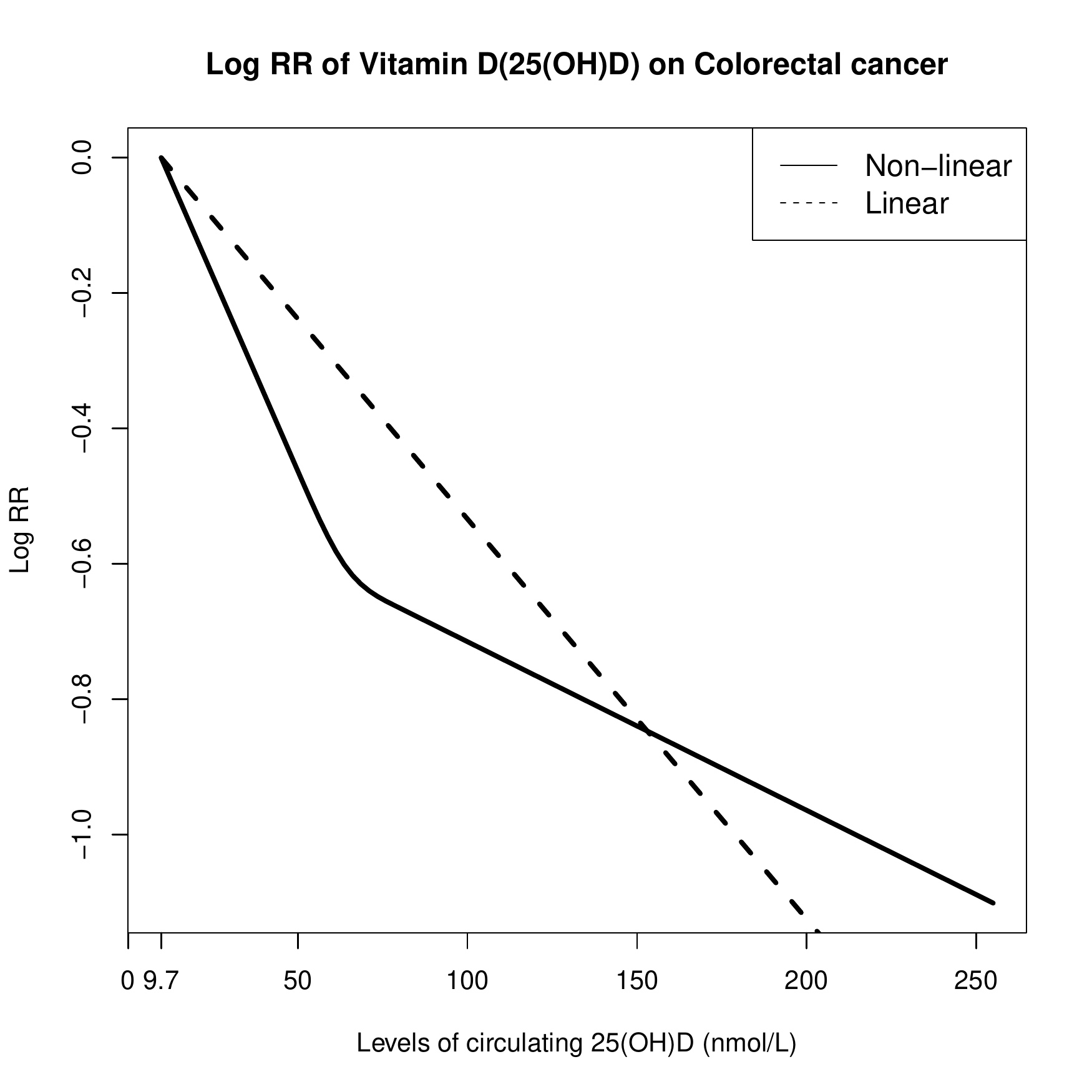}
	\caption{ Log colorectal cancer RR for levels of circulating 25(OH)D compared to the reference level, 9.734 nmol/L based on the full calibration method.}
	\label{applied}
\end{figure}

In figure \ref{applied}, we plotted the log RR of colorectal cancer on circulating 25(OH)D measurements under both models with and without the nonlinear spline term. We set the reference level to be individuals with minimum 25(OH)D measurement 9.734 nmol/L in the aggregated study.

\section{Discussion}

In this paper, we proposed statistical methods for analyzing pooled nested case-control studies. Our model can estimate a possible nonlinear spline dose-response curve between biomarker measurements and the diseases, which can help evaluate whether the relationship is linear or not. We follow the common practice for study-specific calibration models in which only controls are selected for re-assaying in the reference laboratory. We derived an analytic expression for the variance-covariance matrix of the estimated coefficients in the conditional logistic regression model that takes into account of the uncertainty from fitting the calibration model.

Several remarks and recommendations can be drawn from our work. The full calibration method is preferred compared with the internalized calibration method and naive pooling of the uncalibrated data. The full calibration method has very small average relative bias in all simulation scenarios, and has coverage rate close to 95\% nominal level. As the calibration proportion increases, the internalized calibration method becomes more biased than the full calibration method. Since the calibration model is fitted on controls only, estimates of the model parameters are slightly biased. The bias in the intercept is cancelled out in the approximate likelihood function for full calibration, but not with internalized calibration. 

R code for pooling nested case-control study using restricted spline functions is available at \url{https://www.hsph.harvard.edu/molin-wang/software}.

\section*{Acknowledgements}
We are thankful to 
Circulating Biomarkers and Breast and Colorectal Cancer Consortium team (R01CA152071, PI: Stephanie Smith-Warner; Intramural Research Program, Division of Cancer Epidemiology and Genetics, National Cancer Institute: Regina Ziegler) for conducting the calibration study in the vitamin D example. 
Molin Wang was supported in part by NIH/NCI grant R03CA212799. This work was supported in part by U01CA167552, UM1CA167552, UM1CA186107, and the Intramural Program of the National Cancer Institute, Division of Cancer Epidemiology and Genetics. \textit{Conflict of Interest:} None declared.

\begin{supplement}
\sname{Supplement A}\label{suppA}
\stitle{Appendix and supplementary figures and tables}
\slink[url]{http://www.e-publications.org/ims/support/dowload/imsart-ims.zip}
\sdescription{Supplementary material is available online. Section 1 derives the approxiamte likelihood; Section 2 gives the estimating equations for parameters under control-only calibration study; Section 3 derives the sandwich variance estimator for the parameters; Section 4 gives a mathematical justification for the relative performance of full and internalized calibration model; Section 5 includes addtional figures and tables. }

\end{supplement}

\begin{supplement}
\sname{Supplement B}\label{suppB}
\stitle{R code}
\slink[url]{http://www.e-publications.org/ims/support/dowload/imsart-ims.zip}
\sdescription{R functions for implementing the proposed methods}
\end{supplement}

\begin{sidewaystable}
	\setlength{\tabcolsep}{3.5pt}
	\begin{center}
		\begin{tiny}
			\caption{Comparison of operating characteristics for $\boldsymbol{\beta}_X$ under the model $\text{logit}\left(Y_{sji}=1|X_{sji} \right)=\beta_{0sj}+\beta_{X_1}f_1(X_{sji})+\beta_{X_2}f_2(X_{sji})$ for Internalized calibration(IN), Full calibration(FC) and Naive methods. Relative bias is computed using $\frac{\widehat{\beta}-\beta}{\beta}$, and the reported value in the table is the average over the 1000 simulation replicates. Coverage rate is the proportion of simulations that yield a 95\% confidence interval covering the true parameter. Standard deviation is the square root of the empirical variance of parameter estimates over all replicates; we report $10^3$ times the standard deviation. The calibration proportion(denoted as Calib. size in the table) were set to be 5\%, 15\% and 30\%. $\beta_{X_2}$ is fixed at 0.08. }\label{point1}
			\begin{tabular}{@{}cccccccccccccc@{}} \hline\hline
				\multicolumn{2}{c}{ }&\multicolumn{3}{c}{Relative bias of $\beta_{X_1}$ (SD)}&\multicolumn{3}{c}{Coverage Rate of $\beta_{X_1}$}&\multicolumn{3}{c}{Relative bias of $\beta_{X_2}$ (SD)}&\multicolumn{3}{c}{Coverage Rate of $\beta_{X_3}$}\\ \cmidrule(lr){3-5}\cmidrule(lr){6-8}\cmidrule(lr){9-11}\cmidrule(lr){12-14}
				Calib. size&$\beta_{X_1}$&$\beta_{IN}$&$\beta_{FC}$&$\beta_{N}$&$\beta_{IN}$&$\beta_{FC}$&$\beta_{N}$&$\beta_{IN}$&$\beta_{FC}$&$\beta_{N}$&$\beta_{IN}$&$\beta_{FC}$&$\beta_{N}$\\ \hline\hline
				5\%&$-\log(1.25)$&-0.016(2.621)&-0.001(2.959)&-0.444(0.110)&0.970&0.972&0.458&-0.036(0.192)&-0.006(0.196)&-0.722(0.018)&0.968&0.971&0.222\\
				&$-\log(1.5)$&-0.012(2.601)&-0.003(2.972)&-0.237(0.160)&0.964&0.966&0.518&-0.052(0.220)&-0.021(0.228)&-0.360(0.025)
				&0.962&0.962&0.736\\
				&$-\log(1.75)$&-0.011(2.846)&-0.004(3.288)&-0.166(0.207)&0.964&0.966&0.569&-0.085(0.258)&-0.053(0.257)&-0.076(0.033)&0.957&0.959&0.941\\
				&$-\log(2)$&-0.014(2.828)&-0.008(3.212)&-0.128(0.261)&0.968&0.970&0.642&-0.119(0.337)&-0.087(0.339)&0.182(0.040)&0.955&0.958&0.888\\
				&$-\log(2.25)$&-0.010(3.707)&-0.005(4.481)&-0.100(0.324)&0.975&0.980&0.707&-0.108(0.357)&-0.075(0.371)&0.411(0.051)&0.954&0.955&0.766\\
				&$-\log(2.5)$&-0.011(3.785)&-0.006(4.486)&-0.082(0.405)&0.961&0.963&0.754&-0.120(0.385)&-0.085(0.393)&0.620(0.066)&0.944&0.944&0.609\\
				&$-\log(2.75)$&-0.010(4.444)&-0.006(5.136)&-0.072(0.452)&0.965&0.963&0.788&-0.120(0.517)&-0.087(0.535)&0.780(0.075)&0.954&0.954&0.497\\\hline
				15\%&$-\log(1.25)$&-0.071(0.694)&-0.023(0.874)&-0.441(0.120)&0.966&0.969&0.476&-0.136(0.095)&-0.043(0.096)&-0.708(0.019)&0.944&0.953&0.267\\
				&$-\log(1.5)$&-0.039(0.854)&-0.011(1.134)&-0.242(0.154)&0.957&0.956&0.511&-0.134(0.115)&-0.038(0.115)&-0.364(0.024)
				&0.947&0.952&0.739\\
				&$-\log(1.75)$&-0.024(0.836)&-0.003(1.123)&-0.171(0.197)&0.955&0.962&0.564&-0.128(0.145)&-0.029(0.147)&-0.088(0.031)&0.941&0.950&0.937\\
				&$-\log(2)$&-0.018(0.911)&0.000(1.243)&-0.127(0.265)&0.961&0.966&0.638&-0.119(0.174)&-0.019(0.175)&0.194(0.043)&0.951&0.949&0.897\\
				&$-\log(2.25)$&-0.014(1.218)&0.003(1.797)&-0.097(0.326)&0.961&0.972&0.722&-0.149(0.215)&-0.046(0.217)&0.421(0.051)&0.944&0.947&0.765\\
				&$-\log(2.5)$&-0.021(1.108)&-0.006(1.595)&-0.083(0.395)&0.941&0.948&0.761&-0.184(0.247)&-0.076(0.249)&0.620(0.064)&0.938&0.950&0.620\\
				&$-\log(2.75)$&-0.014(1.273)&0.001(1.926)&-0.067(0.461)&0.948&0.960&0.805&-0.169(0.295)&-0.061(0.299)&0.797(0.076)&0.941&0.950&0.481\\\hline
				30\%&$-\log(1.25)$&-0.099(0.519)&-0.003(0.455)&-0.433(0.113)&0.943&0.955&0.477&-0.211(0.083)&-0.023(0.084)&-0.708(0.019)&0.937&0.954&0.265   \\
				&$-\log(1.5)$&-0.051(0.514)&0.006(0.561)&-0.232(0.164)&0.944&0.951&0.533&-0.197(0.096)&-0.005(0.096)&-0.359(0.026)&0.948&0.957&0.751\\
				&$-\log(1.75)$&-0.045(0.564)&-0.002(0.659)&-0.167(0.211)&0.940&0.960&0.576&-0.236(0.129)&-0.041(0.129)&-0.077(0.034)&0.941&0.962&0.926\\
				&$-\log(2)$&-0.040(0.581)&-0.003(0.755)&-0.128(0.258)&0.937&0.961&0.631&-0.255(0.156)&-0.054(0.155)&0.187(0.042)&0.936&0.956&0.900\\
				&$-\log(2.25)$&-0.036(0.661)&-0.003(0.885)&-0.101(0.328)&0.926&0.931&0.698&-0.281(0.193)&-0.074(0.193)&0.411(0.052)&0.922&0.933&0.769\\
				&$-\log(2.5)$&-0.031(0.719)&0.001(0.988)&-0.083(0.409)&0.939&0.954&0.749&-0.242(0.233)&-0.030(0.234)&0.621(0.065)&0.938&0.948&0.613\\
				&$-\log(2.75)$&-0.035(0.835)&-0.006(1.109)&-0.072(0.459)&0.926&0.954&0.794&-0.301(0.274)&-0.089(0.272)&0.786(0.076)&0.910&0.943&0.506\\
				\hline\hline				
			\end{tabular}
		\end{tiny}
	\end{center}
\end{sidewaystable}

\begin{sidewaystable}
	\setlength{\tabcolsep}{3.5pt}
	\begin{center}
		\begin{tiny}
			\caption{Comparison of operating characteristics for $\boldsymbol{\beta}_X$ under the model $\text{logit}\left(Y_{sji}=1|X_{sji} \right)=\beta_{0sj}+\beta_{X_1}f_1(X_{sji})+\beta_{X_2}f_2(X_{sji})$ for Internalized calibration(IN), Full calibration(FC) and Naive methods. Relative bias is computed by $\frac{\widehat{\beta}-\beta}{\beta}$, and the reported value is the average over the 1000 simulation replicates. Coverage rate is the proportion of simulations that yield a 95\% confidence interval covering the true parameter. Standard deviation is the square root of the empirical variance of parameter estimates over all replicates; we report $10^3$ times the standard deviation. The calibration proportion(denoted as Calib. size in the table) were set to be 5\%, 15\% and 30\%. $\beta_{X_1}$ is fixed at $-\log(1.5)\approx -0.41$.}\label{point2}
			\begin{tabular}{@{}cccccccccccccc@{}} \hline\hline
				\multicolumn{2}{c}{ }&\multicolumn{3}{c}{Relative bias of $\beta_{X_1}$ (SD)}&\multicolumn{3}{c}{Coverage Rate of $\beta_{X_1}$}&\multicolumn{3}{c}{Relative bias of $\beta_{X_2}$ (SD)}&\multicolumn{3}{c}{Coverage Rate of $\beta_{X_3}$}\\ \cmidrule(lr){3-5}\cmidrule(lr){6-8}\cmidrule(lr){9-11}\cmidrule(lr){12-14}
				Calib. size&$\beta_{X_2}$&$\beta_{IN}$&$\beta_{FC}$&$\beta_{N}$&$\beta_{IN}$&$\beta_{FC}$&$\beta_{N}$&$\beta_{IN}$&$\beta_{FC}$&$\beta_{N}$&$\beta_{IN}$&$\beta_{FC}$&$\beta_{N}$\\ \hline\hline
				5\%&0.02&-0.013(2.838)&-0.004(3.140)&-0.065(0.178)&0.968&0.970&0.898&-0.216(0.215)&-0.093(0.218)&1.888(0.029)&0.962&0.958&0.624\\
				&0.06&-0.022(3.199)&-0.012(3.528)&-0.181(0.159)&0.960&0.967&0.686&-0.129(0.234)&-0.088(0.247)&-0.124(0.025)&0.949&0.954&0.938\\
				&0.10&-0.025(2.515)&-0.016(2.956)&-0.295(0.144)&0.968&0.969&0.339&-0.081(0.197)&-0.056(0.206)&-0.508(0.023)&0.959&0.957&0.399\\
				&0.14&-0.015(2.815)&-0.006(3.179)&-0.417(0.131)&0.967&0.970&0.078&-0.041(0.194)&-0.023(0.198)&-0.693(0.021)&0.953&0.951&0.010\\
				&0.18&-0.021(2.636)&-0.012(3.149)&-0.528(0.127)&0.968&0.969&0.012&-0.050(0.219)&-0.036(0.225)&-0.780(0.020)&0.950&0.954&0.000\\\hline
				15\%&0.02&-0.027(0.889)&0.001(1.072)&-0.058(0.176)&0.951&0.953&0.922&-0.497(0.126)&-0.117(0.126)&1.919(0.027)&0.928&0.938&0.630\\
				&0.06&-0.033(0.771)&-0.005(0.994)&-0.177(0.161)&0.954&0.970&0.699&-0.190(0.122)&-0.062(0.123)&-0.116(0.026)&0.938&0.948&0.933\\
				&0.10&-0.033(0.867)&-0.005(0.982)&-0.298(0.143)&0.942&0.944&0.335&-0.103(0.114)&-0.026(0.113)&-0.521(0.023)&0.933&0.948&0.376\\
				&0.14&-0.037(0.790)&-0.009(1.059)&-0.418(0.136)&0.952&0.961&0.083&-0.079(0.113)&-0.023(0.114)&-0.692(0.022)&0.954&0.962&0.008\\
				&0.18&-0.043(0.851)&-0.014(1.163)&-0.536(0.118)&0.959&0.964&0.003&-0.072(0.108)&-0.027(0.109)&-0.785(0.019)&0.956&0.958&0.000\\\hline
				30\%&0.02&-0.060(0.546)&-0.005(0.569)&-0.065(0.175)&0.945&0.963&0.916&-0.856(0.110)&-0.104(0.111)&1.897(0.028)&0.932&0.957&0.640\\
				&0.06&-0.056(0.556)&0.000(0.583)&-0.180(0.162)&0.939&0.955&0.682&-0.285(0.101)&-0.029(0.101)&-0.123(0.026)&0.926&0.952&0.928\\
				&0.10&-0.061(0.536)&-0.004(0.619)&-0.294(0.151)&0.927&0.943&0.342&-0.192(0.097)&-0.036(0.097)&-0.517(0.024)&0.929&0.942&0.369\\
				&0.14&-0.064(0.458)&-0.008(0.528)&-0.411(0.135)&0.951&0.954&0.080&-0.139(0.092)&-0.027(0.091)&-0.682(0.022)&0.927&0.953&0.011\\
				&0.l8&-0.069(0.493)&-0.012(0.618)&-0.536(0.125)&0.935&0.957&0.006&-0.118(0.090)&-0.028(0.091)&-0.786(0.020)&0.920&0.950&0.000\\
				\hline\hline				
			\end{tabular}
		\end{tiny}
	\end{center}
\end{sidewaystable}

\begin{sidewaystable}
	\setlength{\tabcolsep}{3.5pt}
	\begin{center}
		\begin{tiny}
			\caption{Comparison of operating characteristics for $\boldsymbol{\beta}_X$ under the model $\text{logit}\left(Y_{sji}=1|X_{sji} \right)=\beta_{0sj}+\beta_{X_1}f_1(X_{sji})+\beta_{X_2}f_2(X_{sji})$ for Internalized calibration(IN), Full calibration(FC) and Naive methods with different $\frac{\sigma_{ws}^2}{\sigma_{X}^2}$. Relative bias is computed using $\frac{\widehat{\beta}-\beta}{\beta}$, and the reported value is the average over the 1000 simulation replicates. Coverage rate is the proportion of simulations that yield a 95\% confidence interval covering the true parameter. Standard deviation is the square root of the empirical variance of parameter estimates over all replicates; we report $10^3$ times the standard deviation. The calibration proportion(denoted as Calib. size in the table) were set to be 5\%, 15\% and 30\%. $\beta_{X_1}=-0.25,\beta_{X_2}=0.08$.}\label{sense2}
			\begin{tabular}{@{}cccccccccccccc@{}} \hline\hline
				\multicolumn{2}{c}{ }&\multicolumn{3}{c}{Relative bias of $\beta_{X_1}$ (SD)}&\multicolumn{3}{c}{Coverage Rate of $\beta_{X_1}$}&\multicolumn{3}{c}{Relative bias of $\beta_{X_2}$ (SD)}&\multicolumn{3}{c}{Coverage Rate of $\beta_{X_3}$}\\ \cmidrule(lr){3-5}\cmidrule(lr){6-8}\cmidrule(lr){9-11}\cmidrule(lr){12-14}
				Calib size&$\frac{\sigma_{ws}^2}{\sigma_x^2}$&$\beta_{IN}$&$\beta_{FC}$&$\beta_{N}$&$\beta_{IN}$&$\beta_{FC}$&$\beta_{N}$&$\beta_{IN}$&$\beta_{FC}$&$\beta_{N}$&$\beta_{IN}$&$\beta_{FC}$&$\beta_{N}$\\ \hline\hline
				5\%&0.75&-0.141(22.094)&-0.051(27.250)&-0.406(0.145)&0.970&0.983&0.509&-0.350(1.966)&-0.152(2.444)&-0.695(0.023)&0.942&0.968&0.375\\
				&0.85&-0.090(8.468)&-0.044(9.954)&-0.409(0.128)&0.966&0.976&0.479&-0.205(0.700)&-0.104(0.800)&-0.682(0.020)&0.944&0.958&0.364\\
				&0.90&-0.053(4.876)&-0.025(5.373)&-0.405(0.125)&0.967&0.976&0.460&-0.106(0.316)&-0.044(0.336)&-0.670(0.020)&0.943&0.954&0.338\\
				&0.95&-0.015(2.398)&-0.002(2.626)&-0.380(0.123)&0.950&0.953&0.504&-0.047(0.184)&-0.017(0.187)&-0.647(0.019)&0.947&0.953&0.352\\
				\hline
				15\%&0.75&-0.311(5.298)&-0.039(6.133)&-0.407(0.151)&0.908&0.976&0.528&-0.713(0.392)&-0.117(0.420)&-0.689(0.024)&0.808&0.957&0.383\\
				&0.85&-0.156(2.250)&-0.018(2.826)&-0.398(0.135)&0.954&0.975&0.500&-0.354(0.200)&-0.050(0.213)&-0.670(0.021)&0.927&0.966&0.344\\
				&0.90&-0.104(1.456)&-0.017(1.794)&-0.387(0.124)&0.951&0.960&0.504&-0.246(0.145)&-0.054(0.152)&-0.655(0.020)&0.933&0.960&0.351\\
				&0.95&-0.045(0.732)&-0.005(0.838)&-0.386(0.130)&0.961&0.961&0.476&-0.114(0.097)&-0.025(0.098)&-0.656(0.020)&0.949&0.960&0.318\\
				\hline
				30\%&0.75&-0.580(3.127)&-0.044(3.189)&-0.403(0.144)&0.639&0.977&0.547&-1.308(0.209)&-0.134(0.224)&-0.683(0.023)&0.390&0.951&0.379\\
				&0.85&-0.316(1.467)&-0.040(1.350)&-0.401(0.132)&0.837&0.967&0.494&-0.705(0.128)&-0.098(0.131)&-0.673(0.021)&0.780&0.951&0.361\\
				&0.90&-0.189(0.913)&-0.017(1.008)&-0.391(0.127)&0.91&0.961&0.478&-0.428(0.105)&-0.049(0.108)&-0.664(0.020)&0.891&0.957&0.336\\
				&0.95&-0.095(0.475)&-0.014(0.456)&-0.377(0.122)&0.924&0.935&0.492&-0.229(0.083)&-0.051(0.084)&-0.640(0.019)&0.920&0.935&0.336\\
				\hline\hline				
			\end{tabular}
		\end{tiny}
	\end{center}
\end{sidewaystable}
\newpage
\begin{table}[ht]
	\setlength{\tabcolsep}{3.5pt}
	\begin{center}
		\caption{Number of cases and controls, size of the calibration study($n_{cal}$), and the estimated intercept and slope of the calibration model for each study in the pooled analysis. }\label{point6}
		\begin{tabular}{@{}ccccc@{}} \hline\hline
			&\multicolumn{4}{c}{Calibration Model}\\ \cmidrule(){2-5}
			Study&Cases/Controls&$n_{cal}$&$\widehat{a}$ (SE)&$\widehat{b}$ (SE)\\\hline
			NHS&348/694&29&-3.56(2.72)&1.13(0.97)\\
			HPFS&267/519&29&3.38(2.95)&0.05(0.04)\\
			\hline\hline				
		\end{tabular}
		\label{calibmodel}
	\end{center}
\end{table}

\begin{table}[ht]
	\centering
	\caption{Point estimates and 95\% confidence intervals for the nonlinear(and linear) association of circulating 25(OH)D and colorectal cancer after adjusting for BMI(overweight or not), physical activity(continuous), smoking(never/ever),Family history of colorectal cancer(yes/no).  }
	\label{clogit}
	\begin{tabular}{ccc}\hline\hline
		Method &$\beta_{X_1}$&$\beta_{X_2}  $\\\hline
		Internalized calibration  & -0.0116 (-0.0214, -0.0017)&7.9307e-06 (-4.6375e-06, 2.0499e-05)\\
		Full calibration&-0.0115 (-0.0213, -0.0017)&7.9885e-06 (-4.8290e-06, 2.0806e-05)\\
		Linear model &-0.0059 (-0.0108, -0.0010)&-\\
		\hline\hline
	\end{tabular}
\end{table}

\newpage

\end{document}